\PassOptionsToPackage{table}{xcolor}
\documentclass[
 reprint,
 superscriptaddress,
 nofootinbib,
 amsmath,amssymb,
 aps,
 pra,
]{revtex4-2}
 
\usepackage{amsmath,amssymb,amsfonts,amsthm,mathtools}
\usepackage{bm}
\usepackage{braket}
\usepackage{dsfont}
\usepackage{mathbbol}
\usepackage{graphicx}
\usepackage{xcolor}
\usepackage{hyperref}
\usepackage{pifont}
 
\definecolor{darkred}{rgb}{0.5,0,0}
\definecolor{darkblue}{rgb}{0,0,0.5}
\definecolor{darkgreen}{rgb}{0,0.5,0}
\hypersetup{colorlinks=true,urlcolor=blue,linkcolor=darkblue,citecolor=darkgreen,filecolor=darkred}
 
\theoremstyle{definition}
\newtheorem{lemma}{Lemma}
\newtheorem{theorem}{Theorem}
\newtheorem{definition}{Definition}
\newtheorem{proposition}{Proposition}
\newtheorem{corollary}{Corollary}
\newtheorem*{remark}{Remark}
 
\newcommand{\conv}{\text{conv}}
\newcommand{\mbb}{\mathbb}
\newcommand{\mc}{\mathcal}
\newcommand{\msf}{\mathsf}
\newcommand{\tr}{\text{Tr}}
\newcommand{\Herm}{\text{Herm}}

\newcommand{\Sep}{\text{Sep}}
\newcommand{\norm}[1]{\left\lVert#1\right\rVert}
\newcommand{\proj}[1]{\lvert #1\rangle \langle #1\rvert}

\newcounter{protocol}
\newenvironment{protocol}[1]{%
  \par\addvspace{0.8\baselineskip}\noindent\begin{minipage}{\linewidth}%
  \refstepcounter{protocol}\hrule\vspace{0.4em}%
  \noindent\textbf{Protocol \theprotocol. #1}\par\smallskip
}{%
  \par\vspace{0.4em}\hrule\end{minipage}\par\addvspace{0.8\baselineskip}
}
\newenvironment{protocolsteps}{%
  \begin{enumerate}%
  \setlength{\itemsep}{0.2em}\setlength{\parsep}{0pt}\setlength{\parskip}{0pt}%
}{\end{enumerate}}
 
\begin{document}
 
\title{Complexity of self-consistent entanglement certification}
 
\author{Yujie Zhang}
\email{yujie4physics@gmail.com}
\affiliation{Institute for Quantum Computing and Department of Physics \& Astronomy, University of Waterloo, 200 University Ave W, Waterloo, Ontario, N2L 3G1, Canada}
\affiliation{Perimeter Institute for Theoretical Physics, 31 Caroline Street North, Waterloo, Ontario, Canada N2L 2Y5}

\begin{abstract}
In recent work~\cite{ZhangEntCert2025}, we proposed a self-consistent approach to entanglement certification based on generalized noncontextuality. It requires no prior characterization of the measurement devices and, given access to all local measurements, can certify every entangled state in a Bell circuit. Here, we study the complexity of this protocol in a finite experiment. At fixed local dimension $d$, we show that the optimal number of distinct local effects needed to reach a trace-distance \textit{entanglement resolution} $\gamma$ is $\Theta_d(\gamma^{-(d-1)})$. The same scaling is necessary even when the measurement design may be tailored to the known target state. Independent Haar-random projective measurements instead require $\Theta_d(\gamma^{-(d-1)}\log(1/\gamma))$ effects. Finally, we show that, once the exact operational identities are known, $\Theta_d(\gamma^{-2}\log(1/\delta))$ copies of the target state are necessary and sufficient for certification with error probability at most $\delta$. Determining the optimal sample complexity when those identities must instead be inferred from the same finite target-state data remains open.
\end{abstract}
 
\maketitle
 \section{Introduction}
 
Certifying whether an experimentally prepared bipartite state is entangled is a central task in quantum information. Existing approaches differ in the assumptions they make about the measurement devices and in the range of entangled states they can certify. Entanglement witnesses and quantum state tomography can, in principle, certify any entangled state, but they require a prior characterization of the measurements~\cite{Horodecki2009,GuhneToth2009}. Bell inequalities remove this requirement, but they certify only those entangled states that are Bell nonlocal~\cite{Werner1989,Barrett2002,Brunner2014}. Steering inequalities retain a trusted description of one party and thus interpolate between these two cases, but they cannot certify unsteerable entangled states~\cite{Wiseman2007}. Other approaches that certify all entangled states require either trusted quantum inputs~\cite{Buscemi2012,Branciard2013}, or a causal structure beyond the Bell circuit together with additional entangled sources~\cite{Bowles2018}.
 
In recent work~\cite{ZhangEntCert2025}, we proposed a new approach to entanglement certification based on generalized noncontextuality. Such a protocol uses only the Bell circuit and requires no prior characterization of the measurements, yet can in principle certify every entangled state. A generalized noncontextuality test asks whether the observed statistics admit a \emph{noncontextual ontological model} (NCOM), i.e., an ontological model that preserves the causal structure and the \emph{operational identities} satisfied by the experimental procedures~\cite{Spekkens2005,SchmidSpekkensWolfe2018,SchmidSelbyPuseySpekkens2024}. For entanglement certification, the relevant identities are those among each party's measurement effects in the Bell circuit. Separable states always yield statistics that are NCOM-realizable relative to these identities, and, when all local measurements are available, only separable states do. Under a tomographic completeness condition, these operational identities can be inferred from the observed statistics by self-consistent tomography~\cite{Nielsen2021,Mazurek2021}, and both the identities and the entanglement certification verdict are independent of the tomographic gauge.

The universality of the self-consistent entanglement-certification protocol relies on access to all local measurements~\cite{ZhangEntCert2025}. This leaves open whether finitely many measurements suffice to certify every entangled state and, if not, what a finite set of measurements can achieve. This question matters in practice, since no experiment can implement all local measurements. To make it quantitative, we fix the local dimension $d$ and ask how many distinct local effects, counted up to positive rescaling, are needed to certify every state whose trace-distance entanglement is at least $\gamma$. We call this number the \emph{effect complexity}. The answer to the first question is negative: no finite set of measurements certifies every entangled state. Our main result quantifies what a finite set can achieve: at fixed $d$, the optimal effect complexity is $\Theta_d(\gamma^{-(d-1)})$. This scaling is attained by unions of rank-one projective measurements, and it remains necessary even when the measurements are tailored to a known target state.
 
The paper proceeds as follows. After reviewing the protocol, we define in Sec.~\ref{sec:resolution-main} the \emph{covering infidelity} $\epsilon$ of a family of local effects, namely the largest infidelity between a pure state and the closest normalized effect. We show that, once $\epsilon$ is small enough, the largest trace-distance entanglement that the family leaves uncertified scales linearly with $\epsilon$, with dimension-dependent constants. The lower bound is witnessed by Werner states, which depend only on $\epsilon$ and not on the measurements. In Sec.~\ref{sec:effect-complexity-main}, we combine this with standard $\epsilon$-net covering arguments~\cite{Szarek1998} to obtain the optimal effect complexity $\Theta_d(\gamma^{-(d-1)})$, while independent Haar-random projective measurements require $\Theta_d(\gamma^{-(d-1)}\log(1/\gamma))$ effects, reflecting the logarithmic overhead of random coverings~\cite{Janson1986,ReznikovSaff2016}. In Sec.~\ref{sec:sampling-main}, we show that, if the exact operational identities are supplied independently of the target data, $\Theta_d(\gamma^{-2}\log(1/\delta))$ copies of the target state are necessary and sufficient for certification with error probability at most $\delta$. Finally, we discuss the fully self-consistent setting, in which the operational identities are inferred from the same data as the target statistics. The statistical uncertainty in the identities must then enter the certification protocol, and determining the resulting sample complexity remains open.
 
\section{Preliminaries}
 
Throughout, $\dim\mc H^A=\dim\mc H^B=d\geq2$ and $\mc D_{AB}:=\mc D(\mc H^A\otimes\mc H^B)$. Following Ref.~\cite{ZhangEntCert2025}, we consider a bipartite Bell circuit $\mc B=(\msf N,\msf M,\msf P)$, where $\msf N=\{\{N^A_{a|x}\}_a\}_x$ and $\msf M=\{\{M^B_{b|y}\}_b\}_y$ are Alice's and Bob's local multi-measurements, each with finitely many settings and outcomes, and $\msf P=\rho^{AB}\in\mc D_{AB}$ is the bipartite state. The operational statistics of the Bell circuit $\mc B$ are
\begin{equation}
p(ab|xy)=
\tr[(N^A_{a|x}\otimes M^B_{b|y})\rho^{AB}].
\label{eq:finite-statistics}
\end{equation}
 
As shown in Ref.~\cite{ZhangEntCert2025}, the entanglement of $\rho^{AB}$ can be certified by testing whether these statistics are NCOM-realizable relative to $\mc O_{\rm all}(\msf N)\cup\mc O_{\rm all}(\msf M)$, i.e., if $p(ab|xy)$ admits a decomposition
\begin{equation}
p(ab|xy)=\sum_{\lambda_A,\lambda_B}p(\lambda_A,\lambda_B)p(a|x,\lambda_A)p(b|y,\lambda_B),
\label{eq:finite-nc-model}
\end{equation}
in which the local response functions are nonnegative and normalized, and preserve all operational identities among the corresponding effects. That is, letting
\begin{equation}
\mc O_{\rm all}(\msf N):=\left\{\vec\alpha: \sum_{a,x}\alpha_{a,x}N^A_{a|x}=\mbb 0^A
\right\},
\end{equation}
Alice's response functions must satisfy $p(a|x,\lambda_A)\ge 0$, $\sum_a p(a|x,\lambda_A)=1$, and
\begin{equation}
\sum_{a,x}\alpha_{a,x}p(a|x,\lambda_A)=0\quad
\forall\vec\alpha\in\mc O_{\rm all}(\msf N),
\label{eq:response-polytope-A}
\end{equation}
and analogously for Bob. Let $\mbb P_{\msf N}$ and $\mbb P_{\msf M}$ denote the resulting noncontextual measurement-assignment polytopes, i.e., the sets of all admissible $\{p(a|x,\lambda_A)\}_{a,x}$ and $\{p(b|y,\lambda_B)\}_{b,y}$, respectively~\cite{SchmidSpekkensWolfe2018,ZhangEntCert2025,Zhang2026quantifiers}. The NCOM-realizable statistics then form the polytope
\begin{equation}
\mc{NC}_{\rm aattt}(\mc B)=\conv\{\xi^A\otimes\xi^B:\xi^A\in\mbb P_{\msf N},\xi^B\in\mbb P_{\msf M}\},
\label{eq:finite-nc-polytope}
\end{equation}
which is termed the noncontextual polytope and depends on the circuit only through $\mc O_{\rm all}(\msf N)$ and $\mc O_{\rm all}(\msf M)$. Every separable state generates statistics in $\mc{NC}_{\rm aattt}(\mc B)$, so statistics outside this polytope certify entanglement~\cite{ZhangEntCert2025}.
 
The same noncontextuality constraints can be expressed in operator space. A noncontextual response function is linear in the effects, so it can be written as $p(a|x,\lambda_A)=\tr(FN^A_{a|x})$ for a Hermitian operator $F$ with unit trace and nonnegative overlap with every effect of $\msf N$, and similarly for Bob. This leads us to define the sets of local pseudo-states
\begin{subequations}
\begin{align}
\mc K_A&:=\{F=F^\dagger:\tr F=1,\ \tr(FN^A_{a|x})\geq0\ \forall a,x\},
\label{eq:response-body-A}\\
\mc K_B&:=\{G=G^\dagger:\tr G=1,\ \tr(GM^B_{b|y})\geq0\ \forall b,y\},
\label{eq:response-body-B}
\end{align}
\label{eq:response-body}
\end{subequations}
and
\begin{equation}
\mc C_{\msf N,\msf M}:=\mc D_{AB}\cap\conv\{F\otimes G:F\in\mc K_A, G\in\mc K_B\}.
\label{eq:pseudoseparable}
\end{equation}
Every density operator belongs to $\mc K_A$ (respectively $\mc K_B$), so $\Sep\subseteq\mc C_{\msf N,\msf M}$. For a finite family of effects, however, $\mc K_A$ and $\mc K_B$ also contain nonpositive operators, and $\mc C_{\msf N,\msf M}$ may contain entangled states.
 
\begin{lemma}
\label{lem:finite-test}
If $\msf N$ and $\msf M$ are informationally complete, the statistics of $\rho^{AB}$ are NCOM-realizable relative to
$\mc O_{\rm aattt}(\mc B):=\mc O_{\rm all}(\msf N)\cup\mc O_{\rm all}(\msf M)$ if and only if
$\rho^{AB}\in\mc C_{\msf N,\msf M}$.
\end{lemma}
Here, informational completeness means that the effects of $\msf N$ span $\text{Herm}(\mc H^A)$, and similarly for $\msf M$. The proof, given in the Supplemental Material, represents each noncontextual response function as the trace pairing with a unique operator in $\mc K_A$ or $\mc K_B$. The implication $\rho^{AB}\in\mc C_{\msf N,\msf M}\Rightarrow$ NCOM-realizability holds without informational completeness.

\section{Entanglement resolution}
\label{sec:resolution-main}
 
\begin{definition}
\label{def:entanglement-resolution}
For finite local multi-measurements $\msf N$ and $\msf M$, and an
entanglement measure $E$, the \emph{entanglement resolution} is
\begin{equation}
\gamma^E_{\msf N,\msf M}:=\max_{\rho^{AB}\in\mc C_{\msf N,\msf M}}E(\rho^{AB}).
\label{eq:resolution-general}
\end{equation}
\end{definition}
If $\msf N$ and $\msf M$ are informationally complete, Lemma~\ref{lem:finite-test} implies that $\gamma^E_{\msf N,\msf M}$ is the largest amount of entanglement that the test of Ref.~\cite{ZhangEntCert2025} can leave uncertified. In the measurement-full limit, where $\msf N^{\rm full}$ and $\msf M^{\rm full}$ consist of all local measurements, $\mc K_A$ and $\mc K_B$ reduce to the local state spaces by the generalized Gleason theorem~\cite{Busch2003}, so that $\mc C_{\msf N^{\rm full},\msf M^{\rm full}}=\Sep$~\cite{ZhangEntCert2025} and hence $\gamma^E_{\msf N^{\rm full},\msf M^{\rm full}}=0$. In this work, we mainly quantify the resolution using the trace-distance entanglement~\cite{VedralEtAl1997}
\begin{equation}
E_{\rm tr}(\rho^{AB}):=
\min_{\sigma^{AB}\in\Sep}
\frac12\norm{\rho^{AB}-\sigma^{AB}}_1.
\label{eq:etr-def}
\end{equation}
Since $\Sep\subseteq\mc C_{\msf N,\msf M}$, $\gamma^{\rm tr}_{\msf N,\msf M}$ is the Hausdorff distance between $\mc C_{\msf N,\msf M}$ and $\Sep$ with respect to the trace distance $\frac12\norm{\rho-\sigma}_1$, where $\norm{X}_1:=\tr\sqrt{X^\dagger X}$.
 
To quantify this resolution, we associate a geometric parameter with each party's set of effects. Specifically, we define the \emph{covering infidelities} of $\msf N$ and $\msf M$ as
\begin{subequations}
\label{eq:covering-infidelity}
\begin{align}
\epsilon_A&:=\max_{P_u=\proj{u}}\min_{a,x}\left[1-\frac{\tr(P_uN_{a|x}^A)}{\tr N_{a|x}^A}\right],\label{eq:covering-infidelity-A}\\
\epsilon_B&:=\max_{P_u=\proj{u}}\min_{b,y}\left[1-\frac{\tr(P_uM_{b|y}^B)}{\tr M_{b|y}^B}\right],\label{eq:covering-infidelity-B}
\end{align}
\end{subequations}
and we set $\epsilon:=\max\{\epsilon_A,\epsilon_B\}$. In other words, every pure state has normalized overlap at least $1-\epsilon_X$ with some effect of party $X$, and some pure state attains this bound. For a rank-one effect $Q=wP_v$, one has $1-\tr(P_uQ)/\tr Q=D_{\rm tr}(P_u,P_v)^2$, where $D_{\rm tr}(\rho,\sigma):=\frac12\norm{\rho-\sigma}_1$. Hence, for a family of rank-one effects, the normalized effects form a $\sqrt{\epsilon_X}$-net of the pure states in trace distance, which is the standard notion of covering for pure quantum states~\cite{HaydenLeungShorWinter2004,GavinskyIto2013}, and $\sqrt{\epsilon_X}$ is the smallest radius for which this holds. We work with the infidelity $\epsilon_X$ rather than the radius $\sqrt{\epsilon_X}$ because, as we show below, the entanglement resolution scales linearly with $\epsilon$.
 
\begin{theorem}
\label{thm:resolution-covering}
For every $d$, there exist constants $c_d,C_d>0$ such that every pair of finite local multi-measurements with $\epsilon\leq\epsilon_d$ satisfies
\begin{equation}
c_d\epsilon\leq\gamma^{\rm tr}_{\msf N,\msf M}\leq C_d\epsilon.
\label{eq:resolution-covering-bounds}
\end{equation}
Consequently, $\gamma^{\rm tr}_{\msf N,\msf M}=\Theta_d(\epsilon)$ as $\epsilon\to0$.
\end{theorem}
 
The proof in the Supplemental Material gives $c_d=1/[16d(d+1)]$ and $C_d=2d$, which we have not attempted to optimize. Since the setting is symmetric under exchanging Alice and Bob, we assume without loss of generality that $\epsilon=\epsilon_A\geq\epsilon_B$.
 
The upper bound rests on the fact that $\epsilon$ limits how negative a local pseudo-state in Eq.~\eqref{eq:response-body} can be. If $F\in\mc K_A$ and $P_u$ projects onto an eigenvector of its smallest eigenvalue, then nonnegativity of $F$ on the effect that best covers $P_u$ forces $\lambda_{\min}(F)\geq-\epsilon/(1-d\epsilon)$, and the same bound holds for every $G\in\mc K_B$. A local depolarization of strength $d\epsilon$ on each side therefore maps every local pseudo-state to a density operator, and hence every state in $\mc C_{\msf N,\msf M}$ to a separable state at trace distance at most $2d\epsilon$.
 
The lower bound is witnessed by a family of states that does not depend on the measurements. Let $\mbb S$ denote the swap operator, and consider the Werner states~\cite{Werner1989}
\begin{equation}
\rho^{AB}_\gamma:=\frac{(d+2\gamma)\mbb 1^{AB}-(1+2d\gamma)\mbb S^{AB}}{d(d^2-1)},
\label{eq:werner-hard-main}
\end{equation}
which satisfy $E_{\rm tr}(\rho^{AB}_\gamma)=\max\{0,\gamma\}$ for $-1/2\leq\gamma\leq1/2$. In the Supplemental Material, we show that $\rho_\gamma\in\mc C_{\msf N,\msf M}$ for $\gamma=\epsilon/[16d(d+1)]$, and hence, by convexity, for all smaller $\gamma\geq0$. Since $\rho_\gamma$ is invariant under $U\otimes U$, the same state lies in $\mc C_{\msf N,\msf M}$ for every measurement configuration with covering infidelity at least $\epsilon$, whatever its worst-covered direction.
\begin{remark}
The generalized robustness $E_{\rm gr}$~\cite{Steiner2003,Brandao2005} obeys
\begin{equation}
\frac1dE_{\rm gr}(\rho^{AB})\leq E_{\rm tr}(\rho^{AB})\leq\frac{E_{\rm gr}(\rho^{AB})}{1+E_{\rm gr}(\rho^{AB})},
\label{eq:tr-gr-comparison}
\end{equation}
hence $\gamma^{\rm gr}_{\msf N,\msf M}=\Theta_d(\epsilon)$. Since the max-relative entropy of entanglement is $E_{\max}=\log_2(1+E_{\rm gr})$~\cite{Datta2009}, $\gamma^{\max}_{\msf N,\msf M}=\Theta_d(\epsilon)$.
\end{remark}
 
\begin{corollary}
\label{cor:no-finite-universality}
For every pair of finite local multi-measurements $\msf N$ and $\msf M$, $\Sep\subsetneq\mc C_{\msf N,\msf M}$, i.e., some entangled state has NCOM-realizable statistics.
\end{corollary}
Indeed, finitely many effects define finitely many normalized directions, hence $\epsilon\in(0,1]$.
 
\section{Effect complexity}
\label{sec:effect-complexity-main}
 
\subsection{Optimal effect complexity}
 
Theorem~\ref{thm:resolution-covering} reduces the performance of a finite measurement family to the single parameter $\epsilon$, and thereby connects the effect complexity to a covering problem. For rank-one effects, $\sqrt{\epsilon_A}$ is the trace-distance covering radius of the normalized effects in the set of pure states, so achieving a given $\epsilon$ amounts to constructing a $\sqrt{\epsilon}$-net of pure states. Such nets are standard tools in quantum information~\cite{HaydenLeungShorWinter2004,GavinskyIto2013,AubrunSzarek2017}. Since the pure states of a $d$-dimensional system form a manifold of real dimension $2(d-1)$, a $\sqrt{\epsilon}$-net requires of order $\epsilon^{-(d-1)}$ points~\cite{GavinskyIto2013}.
 
To count effects, we call two nonzero effects cone equivalent if $Q=tQ'$ for some $t>0$. Such effects have the same normalized overlaps and impose the same constraint in Eq.~\eqref{eq:response-body}. More generally, cone-equivalent effects are interchangeable in generalized noncontextuality: replacing effects by cone-equivalent ones preserves the existence of a noncontextual model~\cite{accessiblefrags}. We therefore count cone-inequivalent effects,
\begin{equation}
n_A:=\text{Card}_{\rm cone}(\msf N),\qquad n_B:=\text{Card}_{\rm cone}(\msf M),
\label{eq:effect-complexity}
\end{equation}
and define the optimal covering infidelity for at most $n$ local effects as
\begin{equation}
\epsilon_{\rm opt}(n):=\inf_{\msf N}\{\epsilon_A(\msf N):\text{Card}_{\rm cone}(\msf N)\leq n\}.
\label{eq:epsilon-opt-def}
\end{equation}
 
\begin{lemma}
\label{lem:effect-covering}
For fixed $d$,
$\epsilon_{\rm opt}(n)=\Theta_d(n^{-1/(d-1)})$
as $n\to\infty$. The upper bound is attained by a set of rank-one projective measurements.
\end{lemma}
 
For rank-one effects, Lemma~\ref{lem:effect-covering} is the standard covering law for pure states: an optimal $n$-point net has trace-distance covering radius $\Theta_d(n^{-1/[2(d-1)]})$~\cite{Szarek1998,HaydenLeungShorWinter2004}. The additional content of Lemma~\ref{lem:effect-covering} is that higher-rank effects cannot improve this scaling, since in the fine-covering regime the pure states covered by an effect are contained in those covered by the projector onto its eigenvector that has the largest eigenvalue, and that the covering directions can be completed to projective measurements at a cost of at most a factor $d$.

Combining Lemma~\ref{lem:effect-covering} with Theorem~\ref{thm:resolution-covering}, we obtain the effect complexity of universal certification at entanglement resolution $\gamma$. Let $n^{\rm tr}_{\rm opt}(\gamma)$ denote the smallest $n$ for which there exist local multi-measurements with $n_A,n_B\leq n$ and $\gamma^{\rm tr}_{\msf N,\msf M}<\gamma$, i.e., which certify every state with $E_{\rm tr}(\rho^{AB})\geq\gamma$.
 
\begin{theorem}
\label{thm:effect-complexity}
For fixed $d$, $n^{\rm tr}_{\rm opt}(\gamma)=\Theta_d(\gamma^{-(d-1)})$ as $\gamma\to0$.
\end{theorem}

Indeed, Lemma~\ref{lem:effect-covering} gives the optimal covering infidelity $\Theta_d(n^{-1/(d-1)})$ for each party, and Theorem~\ref{thm:resolution-covering} converts this directly into entanglement resolution.
 
 
\subsection{Haar-random measurements}
 
The optimal effect complexity relies on effect directions that form an efficient covering. We now consider instead randomly chosen measurements. Let $\msf N$ consist of the first $m$ elements of a sequence of independent Haar-random rank-one projective measurements, i.e., of independent Haar-random orthonormal bases, so that $\msf N$ contains $n=dm$ effects, and let $\epsilon_{\rm Haar}(n)$ denote its covering infidelity.
 
\begin{lemma}
\label{lem:haar-covering}
For fixed $d\geq2$,
\begin{equation}
\epsilon_{\rm Haar}(n)=\Theta_d \left[\left(\frac{\log n}{n}\right)^{1/(d-1)}\right]
\end{equation}
with probability tending to one as $n\to\infty$.
\end{lemma}
 
Compared with Lemma~\ref{lem:effect-covering}, random directions lose a logarithmic factor. This is the scaling expected for random coverings of a manifold~\cite{Janson1986,ReznikovSaff2016}. The Supplemental Material gives a self-contained proof, first for independent Haar-random directions and then for projective measurements. Applying Lemma~\ref{lem:haar-covering} to both parties together with Theorem~\ref{thm:resolution-covering} gives the following.
 
\begin{proposition}
\label{prop:haar-effect-complexity}
Let $\msf N$ and $\msf M$ be sampled as above, independently for Alice and Bob, and set $\gamma_{\rm Haar}^{\rm tr}(n):=\gamma_{\msf N_m,\msf M_m}^{\rm tr}$ with $n=dm$. For fixed $d\geq2$,
\begin{equation}
\gamma_{\rm Haar}^{\rm tr}(n)=\Theta_d \left[\left(\frac{\log n}{n}\right)^{1/(d-1)}\right]
\label{eq:gamma-haar}
\end{equation}
with probability tending to one as $n\to\infty$. Moreover, for every fixed $p\in(0,1)$, the smallest $n$ for which $\gamma_{\rm Haar}^{\rm tr}(n)<\gamma$ holds with probability at least $p$ is
\begin{equation}
n_{\rm Haar}^{\rm tr}(\gamma,p)=\Theta_d \left[\gamma^{-(d-1)}\log\frac1\gamma\right]
\label{eq:haar-inversion}
\end{equation}
as $\gamma\to0$.
\end{proposition}
 
Independent Haar sampling therefore costs only a logarithmic factor relative to the optimal effect complexity of Theorem~\ref{thm:effect-complexity}. 
 
\section{Sample complexity}
\label{sec:sampling-main}
 
We now count the copies of the target state needed for certification, assuming i.i.d.\ target copies and stationary, memoryless devices. Because of statistical fluctuations, the observed relative frequencies in general do not satisfy the operational identities exactly~\cite{ZhangEntCert2025,SchmidSelbySpekkensObjections2025}, so any statement about a finite number of copies must specify how the identities are obtained. Here, we consider the setting in which the complete and exact identities $\mc O_{\rm all}(\msf N)$ and $\mc O_{\rm all}(\msf M)$ are supplied independently of the target data, for instance by a separate characterization stage using an unbounded number of ancillas, and thus fix $\mc{NC}_{\rm aattt}(\mc B)$ before any target copy is measured. Following the standard formulation of quantum certification~\cite{KlieschRoth2021,MontanaroDeWolf2016,DuEtAl2025}, a protocol returns either $\mathsf{certify}$ or $\mathsf{inconclusive}$, and we require
\begin{subequations}
\begin{align}
&\Pr(\mathsf{certify})\leq\delta\quad\text{whenever }p\in\mc{NC}_{\rm aattt}(\mc B),
\label{eq:sampling-soundness}\\
&\Pr(\mathsf{certify})\geq1-\delta\quad\text{whenever }E_{\rm tr}(\rho^{AB})\geq\gamma.
\label{eq:sampling-completeness}
\end{align}
\end{subequations}
Let $N_{\rm tar}^{(d)}(\gamma,\delta)$ denote the smallest number of target copies for which a single protocol satisfies both, when every setting of the devices is a rank-one projective measurement and $\epsilon\leq\gamma/(16d)$.
 
\begin{theorem}
\label{thm:known-identities-sampling}
For fixed $d$, sufficiently small $\gamma$, and $0<\delta\leq1/4$,
\begin{equation}
N_{\rm tar}^{(d)}(\gamma,\delta)=\Theta_d\left(\gamma^{-2}\log\frac1\delta\right).
\label{eq:known-identities-rate}
\end{equation}
The lower bound holds for all local measurements and every protocol, including those that use joint measurements on all copies.
\end{theorem}
The protocol and proofs are given in the Supplemental Material.
 
Once the identities are supplied, the number of target copies thus does not grow with the number of distinct effects: only $d^4$ product probabilities are sampled on the target, while the remaining effects enter only through the identities. In the protocol of Ref.~\cite{ZhangEntCert2025}, however, the identities are inferred from the same data. The statistical uncertainty of the inferred identities must then be taken into account in the noncontextuality test, and the copies used to infer them must be included in the total number of target copies. Determining the sample complexity in this fully self-consistent setting, and in particular whether the effect complexity then enters the number of copies, remains open.
 
\section{Discussion}
 
In this work, we studied the complexity of the self-consistent entanglement certification protocol of Ref.~\cite{ZhangEntCert2025} when only finitely many local measurements are available. We showed that no finite set of measurements certifies every entangled state, and that the largest trace-distance entanglement that a measurement family leaves uncertified scales linearly with its covering infidelity. At fixed local dimension $d$, the optimal effect complexity at entanglement resolution $\gamma$ is therefore $\Theta_d(\gamma^{-(d-1)})$, whereas independent Haar-random projective measurements require $\Theta_d(\gamma^{-(d-1)}\log(1/\gamma))$ effects. In particular, informational completeness alone does not suffice for universality: nonnegativity on finitely many effects still admits nonpositive local pseudo-states, and the Hausdorff distance between $\mc C_{\msf N,\msf M}$ and $\Sep$ decreases only as the pure states are covered more finely. Repeating a fixed measurement family improves the precision of the statistics but cannot remove its nonzero entanglement resolution, even when the measurements are tailored to a known target state.
 
When the exact operational identities are supplied independently of the target data, we showed that $\Theta_d(\gamma^{-2}\log(1/\delta))$ copies of the target state are necessary and sufficient, independently of the number of distinct effects. This is the same inverse-square dependence on the resolution as for state tomography with calibrated measurements~\cite{HaahEtAl2017,ODonnellWright2016}.
 
The main open problem is the fully self-consistent setting of Ref.~\cite{ZhangEntCert2025}, in which the operational identities are inferred from the same data as the target statistics. A certification statement for a finite number of copies then has to account for the statistical uncertainty of the inferred identities, and the copies used to infer them must be included in the total number of target copies. Reference~\cite{ZhangEntCert2025} discussed two techniques for handling inexact identities in the Bell circuit: the cut-to-your-cloth technique, which infers the identities satisfied by the implemented measurements, and the secondary-procedures technique, which enforces a chosen set of identities by classical post-processing. A finite-copy analysis of these techniques would determine whether, and how, the effect complexity enters the sample complexity of self-consistent certification.

Another natural direction is to study the complexity of self-consistent certification for other types of quantum processes. Generalized noncontextuality also certifies steering in the Bell circuit, using a different subset of operational identities~\cite{ZhangEntCert2025}, and it judges every set of incompatible measurements, every non-entanglement-breaking channel, and every steerable assemblage to be nonclassical~\cite{ZhangReassessing2026}. For each of these tasks, one can ask how many distinct procedures, and how many uses of the target process, are needed for its certification.

A further question concerns the dependence on the local dimension. The constants $c_d$ and $C_d$ in Theorem~\ref{thm:resolution-covering} differ by a factor of order $d^3$, and we do not know whether this gap can be closed.

Finally, the computational complexity of the test remains to be understood. By Lemma~\ref{lem:finite-test} the noncontextuality test decides whether $\rho^{AB}\in\mc C_{\msf N,\msf M}$. The sets $\mc C_{\msf N,\msf M}$ are therefore outer approximations of $\Sep$, at trace-distance Hausdorff distance $\Theta_d(\epsilon)$ from it. Deciding membership in $\Sep$ itself is NP-hard when the local dimension is part of the input~\cite{Gurvits2003,Gharibian2010}. Membership in $\mc C_{\msf N,\msf M}$ is decided by a linear program over $\mc{NC}_{\rm aattt}(\mc B)$, whose vertices are products of vertices of $\mbb P_{\msf N}$ and $\mbb P_{\msf M}$, and whose number of vertices therefore grows only polynomially with the number of effects at fixed $d$~\cite{ZhangEntCert2025,McMullen1970}. With the $O_d(\gamma^{-(d-1)})$ effects of Theorem~\ref{thm:effect-complexity}, the size of this linear program is therefore polynomial in $1/\gamma$. It would be interesting to compare the computational complexity of these outer approximations with other approaches, such as the hierarchy in Ref.~\cite{DohertyParriloSpedalieri2004}.

\section*{Acknowledgments}
Y.Z. thanks Robert Spekkens and David Schmid for helpful discussions and for encouraging the posting of this preliminary draft. Y.Z. acknowledges support from the Institute for Quantum Computing at the University of Waterloo and from Perimeter Institute. Research at the Institute for Quantum Computing is supported by Innovation, Science and Economic Development Canada. Research at Perimeter Institute is supported in part by the Government of Canada through the Department of Innovation, Science and Economic Development, and by the Province of Ontario through the Ministry of Colleges and Universities.

\section*{Statement of AI Usage}
The author formulated the problem of determining the effect complexity of self-consistent entanglement certification, as well as its reduction to a covering-infidelity problem for a set of effects. The optimal effect complexity and the Haar-random complexity were subsequently established through interactions with OpenAI Codex (using models 5.6 and later 6). AI tools were particularly helpful in identifying relevant existing references, improving and streamlining proofs, and extensively exploring the sample complexity of the existing protocol. The author take full responsibility for the mathematical correctness of all results presented in this work.
 
\bibliography{ref}

@article{ZhangEntCert2025,
  author  = {Zhang, Yujie and Spodek, Jonah and Schmid, David and Reid, Carter and Morrison, Liam J. and Jennewein, Thomas and Resch, Kevin J. and Spekkens, Robert W.},
  title   = {Entanglement Certification Using Noncontextuality Inequalities},
  journal = {Phys. Rev. X},
  volume  = {16},
  number  = {3},
  pages   = {031057},
  year    = {2026},
  doi     = {10.1103/dxpr-wp6l},
  url     = {https://doi.org/10.1103/dxpr-wp6l}
}

@article{DuEtAl2025,
  author  = {Du, Zhenyu and Liu, Jinchang and Huber, Elias X. and Liu, Zi-Wen and Ma, Xiongfeng},
  title   = {Certifying localizable quantum properties with constant sample complexity},
  journal = {Phys. Rev. X},
  year    = {2026},
  note    = {Accepted 16 September 2026},
  doi     = {10.1103/rkl6-hfrn},
  url     = {https://doi.org/10.1103/rkl6-hfrn}
}

@article{Horodecki2009,
  author  = {Horodecki, Ryszard and Horodecki, Pawe{\l} and Horodecki, Micha{\l} and Horodecki, Karol},
  title   = {Quantum entanglement},
  journal = {Rev. Mod. Phys.},
  volume  = {81},
  number  = {2},
  pages   = {865},
  year    = {2009},
  doi     = {10.1103/RevModPhys.81.865},
  url     = {https://doi.org/10.1103/RevModPhys.81.865}
}

@article{Werner1989,
  author  = {Werner, Reinhard F.},
  title   = {Quantum states with {E}instein-{P}odolsky-{R}osen correlations admitting a hidden-variable model},
  journal = {Phys. Rev. A},
  volume  = {40},
  number  = {8},
  pages   = {4277},
  year    = {1989},
  doi     = {10.1103/PhysRevA.40.4277},
  url     = {https://doi.org/10.1103/PhysRevA.40.4277}
}

@article{Barrett2002,
  author  = {Barrett, Jonathan},
  title   = {Nonsequential positive-operator-valued measurements on entangled mixed states do not always violate a {B}ell inequality},
  journal = {Phys. Rev. A},
  volume  = {65},
  number  = {4},
  pages   = {042302},
  year    = {2002},
  doi     = {10.1103/PhysRevA.65.042302},
  url     = {https://doi.org/10.1103/PhysRevA.65.042302}
}

@article{Brunner2014,
  author  = {Brunner, Nicolas and Cavalcanti, Daniel and Pironio, Stefano and Scarani, Valerio and Wehner, Stephanie},
  title   = {{B}ell nonlocality},
  journal = {Rev. Mod. Phys.},
  volume  = {86},
  number  = {2},
  pages   = {419},
  year    = {2014},
  doi     = {10.1103/RevModPhys.86.419},
  url     = {https://doi.org/10.1103/RevModPhys.86.419}
}

@article{Wiseman2007,
  author  = {Wiseman, H. M. and Jones, S. J. and Doherty, A. C.},
  title   = {Steering, Entanglement, Nonlocality, and the {E}instein-{P}odolsky-{R}osen Paradox},
  journal = {Phys. Rev. Lett.},
  volume  = {98},
  number  = {14},
  pages   = {140402},
  year    = {2007},
  doi     = {10.1103/PhysRevLett.98.140402},
  url     = {https://doi.org/10.1103/PhysRevLett.98.140402}
}

@article{Buscemi2012,
  author  = {Buscemi, Francesco},
  title   = {All Entangled Quantum States Are Nonlocal},
  journal = {Phys. Rev. Lett.},
  volume  = {108},
  number  = {20},
  pages   = {200401},
  year    = {2012},
  doi     = {10.1103/PhysRevLett.108.200401},
  url     = {https://doi.org/10.1103/PhysRevLett.108.200401}
}

@article{Branciard2013,
  author  = {Branciard, Cyril and Rosset, Denis and Liang, Yeong-Cherng and Gisin, Nicolas},
  title   = {Measurement-Device-Independent Entanglement Witnesses for All Entangled Quantum States},
  journal = {Phys. Rev. Lett.},
  volume  = {110},
  number  = {6},
  pages   = {060405},
  year    = {2013},
  doi     = {10.1103/PhysRevLett.110.060405},
  url     = {https://doi.org/10.1103/PhysRevLett.110.060405}
}

@article{Bowles2018,
  author  = {Bowles, Joseph and {\v{S}}upi{\'c}, Ivan and Cavalcanti, Daniel and Ac{\'i}n, Antonio},
  title   = {Device-Independent Entanglement Certification of All Entangled States},
  journal = {Phys. Rev. Lett.},
  volume  = {121},
  number  = {18},
  pages   = {180503},
  year    = {2018},
  doi     = {10.1103/PhysRevLett.121.180503},
  url     = {https://doi.org/10.1103/PhysRevLett.121.180503}
}

@article{Spekkens2005,
  author  = {Spekkens, R. W.},
  title   = {Contextuality for preparations, transformations, and unsharp measurements},
  journal = {Phys. Rev. A},
  volume  = {71},
  number  = {5},
  pages   = {052108},
  year    = {2005},
  doi     = {10.1103/PhysRevA.71.052108},
  url     = {https://doi.org/10.1103/PhysRevA.71.052108}
}

@article{SchmidSelbySpekkensObjections2025,
  author  = {Schmid, David and Selby, John H. and Spekkens, Robert W.},
  title   = {Addressing some common objections to generalized noncontextuality},
  journal = {Phys. Rev. A},
  volume  = {109},
  number  = {2},
  pages   = {022228},
  year    = {2024},
  doi     = {10.1103/PhysRevA.109.022228},
  url     = {https://doi.org/10.1103/PhysRevA.109.022228}
}

@article{VedralEtAl1997,
  author  = {Vedral, V. and Plenio, M. B. and Rippin, M. A. and Knight, P. L.},
  title   = {Quantifying Entanglement},
  journal = {Phys. Rev. Lett.},
  volume  = {78},
  number  = {12},
  pages   = {2275},
  year    = {1997},
  doi     = {10.1103/PhysRevLett.78.2275},
  url     = {https://doi.org/10.1103/PhysRevLett.78.2275}
}

@article{Steiner2003,
  author  = {Steiner, Michael},
  title   = {Generalized robustness of entanglement},
  journal = {Phys. Rev. A},
  volume  = {67},
  number  = {5},
  pages   = {054305},
  year    = {2003},
  doi     = {10.1103/PhysRevA.67.054305},
  url     = {https://doi.org/10.1103/PhysRevA.67.054305}
}

@article{Brandao2005,
  author  = {Brand{\~a}o, Fernando G. S. L.},
  title   = {Quantifying entanglement with witness operators},
  journal = {Phys. Rev. A},
  volume  = {72},
  number  = {2},
  pages   = {022310},
  year    = {2005},
  doi     = {10.1103/PhysRevA.72.022310},
  url     = {https://doi.org/10.1103/PhysRevA.72.022310}
}

@article{accessiblefrags,
  author  = {Selby, John H. and Schmid, David and Wolfe, Elie and Sainz, Ana Bel{\'e}n and Kunjwal, Ravi and Spekkens, Robert W.},
  title   = {Accessible fragments of generalized probabilistic theories, cone equivalence, and applications to witnessing nonclassicality},
  journal = {Phys. Rev. A},
  volume  = {107},
  number  = {6},
  pages   = {062203},
  year    = {2023},
  doi     = {10.1103/PhysRevA.107.062203},
  url     = {https://doi.org/10.1103/PhysRevA.107.062203}
}

@article{ZyczkowskiSommers2005,
  author  = {{\.Z}yczkowski, Karol and Sommers, Hans-J{\"u}rgen},
  title   = {Average fidelity between random quantum states},
  journal = {Phys. Rev. A},
  volume  = {71},
  number  = {3},
  pages   = {032313},
  year    = {2005},
  doi     = {10.1103/PhysRevA.71.032313},
  url     = {https://doi.org/10.1103/PhysRevA.71.032313}
}

@article{KlieschRoth2021,
  author  = {Kliesch, Martin and Roth, Ingo},
  title   = {Theory of Quantum System Certification},
  journal = {PRX Quantum},
  volume  = {2},
  number  = {1},
  pages   = {010201},
  year    = {2021},
  doi     = {10.1103/PRXQuantum.2.010201},
  url     = {https://doi.org/10.1103/PRXQuantum.2.010201}
}

@article{SchmidSpekkensWolfe2018,
  author  = {Schmid, David and Spekkens, Robert W. and Wolfe, Elie},
  title   = {All the noncontextuality inequalities for arbitrary prepare-and-measure experiments with respect to any fixed set of operational equivalences},
  journal = {Phys. Rev. A},
  volume  = {97},
  number  = {6},
  pages   = {062103},
  year    = {2018},
  doi     = {10.1103/PhysRevA.97.062103},
  url     = {https://doi.org/10.1103/PhysRevA.97.062103}
}

@article{OszmaniecEtAl2024,
  author  = {Oszmaniec, Micha{\l} and Kotowski, Marcin and Horodecki, Micha{\l} and Hunter-Jones, Nicholas},
  title   = {Saturation and Recurrence of Quantum Complexity in Random Local Quantum Dynamics},
  journal = {Phys. Rev. X},
  volume  = {14},
  number  = {4},
  pages   = {041068},
  year    = {2024},
  doi     = {10.1103/PhysRevX.14.041068},
  url     = {https://doi.org/10.1103/PhysRevX.14.041068}
}

@article{Mazurek2021,
  author  = {Mazurek, Michael D. and Pusey, Matthew F. and Resch, Kevin J. and Spekkens, Robert W.},
  title   = {Experimentally Bounding Deviations From Quantum Theory in the Landscape of Generalized Probabilistic Theories},
  journal = {PRX Quantum},
  volume  = {2},
  number  = {2},
  pages   = {020302},
  year    = {2021},
  doi     = {10.1103/PRXQuantum.2.020302},
  url     = {https://doi.org/10.1103/PRXQuantum.2.020302}
}

@article{DohertyParriloSpedalieri2004,
  author  = {Doherty, Andrew C. and Parrilo, Pablo A. and Spedalieri, Federico M.},
  title   = {Complete family of separability criteria},
  journal = {Phys. Rev. A},
  volume  = {69},
  number  = {2},
  pages   = {022308},
  year    = {2004},
  doi     = {10.1103/PhysRevA.69.022308},
  url     = {https://doi.org/10.1103/PhysRevA.69.022308}
}

@article{SchmidSelbyPuseySpekkens2024,
  author  = {Schmid, David and Selby, John H. and Pusey, Matthew F. and Spekkens, Robert W.},
  title   = {A structure theorem for generalized-noncontextual ontological models},
  journal = {Quantum},
  volume  = {8},
  pages   = {1283},
  year    = {2024},
  doi     = {10.22331/q-2024-03-14-1283},
  url     = {https://doi.org/10.22331/q-2024-03-14-1283}
}

@article{Nielsen2021,
  author  = {Nielsen, Erik and Gamble, John King and Rudinger, Kenneth and Scholten, Travis and Young, Kevin and Blume-Kohout, Robin},
  title   = {Gate Set Tomography},
  journal = {Quantum},
  volume  = {5},
  pages   = {557},
  year    = {2021},
  doi     = {10.22331/q-2021-10-05-557},
  url     = {https://doi.org/10.22331/q-2021-10-05-557}
}

@article{HaydenLeungShorWinter2004,
  author  = {Hayden, P. and Leung, Debbie and Shor, Peter W. and Winter, Andreas},
  title   = {Randomizing Quantum States: Constructions and Applications},
  journal = {Commun. Math. Phys.},
  volume  = {250},
  number  = {2},
  pages   = {371--391},
  year    = {2004},
  doi     = {10.1007/s00220-004-1087-6},
  url     = {https://doi.org/10.1007/s00220-004-1087-6}
}

@article{Lindblad1975,
  author  = {Lindblad, G{\"o}ran},
  title   = {Completely positive maps and entropy inequalities},
  journal = {Commun. Math. Phys.},
  volume  = {40},
  number  = {2},
  pages   = {147--151},
  year    = {1975},
  doi     = {10.1007/BF01609396},
  url     = {https://doi.org/10.1007/BF01609396}
}

@article{Janson1986,
  author  = {Janson, Svante},
  title   = {Random coverings in several dimensions},
  journal = {Acta Math.},
  volume  = {156},
  pages   = {83--118},
  year    = {1986},
  doi     = {10.1007/BF02399201},
  url     = {https://doi.org/10.1007/BF02399201}
}

@article{ReznikovSaff2016,
  author  = {Reznikov, A. and Saff, E. B.},
  title   = {The Covering Radius of Randomly Distributed Points on a Manifold},
  journal = {Int. Math. Res. Not.},
  volume  = {2016},
  number  = {19},
  pages   = {6065--6094},
  year    = {2016},
  doi     = {10.1093/imrn/rnv342},
  url     = {https://doi.org/10.1093/imrn/rnv342}
}

@article{Datta2009,
  author  = {Datta, Nilanjana},
  title   = {Max-relative entropy of entanglement, alias log robustness},
  journal = {Int. J. Quantum Inf.},
  volume  = {7},
  number  = {2},
  pages   = {475--491},
  year    = {2009},
  doi     = {10.1142/S0219749909005298},
  url     = {https://doi.org/10.1142/S0219749909005298}
}

@article{Hoeffding1963,
  author  = {Hoeffding, Wassily},
  title   = {Probability Inequalities for Sums of Bounded Random Variables},
  journal = {J. Am. Stat. Assoc.},
  volume  = {58},
  number  = {301},
  pages   = {13--30},
  year    = {1963},
  doi     = {10.1080/01621459.1963.10500830},
  url     = {https://doi.org/10.1080/01621459.1963.10500830}
}

@article{HaahEtAl2017,
  author  = {Haah, Jeongwan and Harrow, Aram W. and Ji, Zhengfeng and Wu, Xiaodi and Yu, Nengkun},
  title   = {Sample-Optimal Tomography of Quantum States},
  journal = {IEEE Trans. Inf. Theory},
  volume  = {63},
  number  = {9},
  pages   = {5628--5641},
  year    = {2017},
  doi     = {10.1109/TIT.2017.2719044},
  url     = {https://doi.org/10.1109/TIT.2017.2719044}
}

@inproceedings{ODonnellWright2016,
  author    = {O'Donnell, Ryan and Wright, John},
  title     = {Efficient quantum tomography},
  booktitle = {Proceedings of the Forty-Eighth Annual {ACM} Symposium on Theory of Computing},
  series    = {STOC '16},
  publisher = {Association for Computing Machinery},
  address   = {New York, NY, USA},
  pages     = {899--912},
  year      = {2016},
  doi       = {10.1145/2897518.2897544},
  url       = {https://doi.org/10.1145/2897518.2897544}
}

@inproceedings{Gurvits2003,
  author    = {Gurvits, Leonid},
  title     = {Classical deterministic complexity of {E}dmonds' Problem and quantum entanglement},
  booktitle = {Proceedings of the Thirty-Fifth Annual {ACM} Symposium on Theory of Computing},
  series    = {STOC '03},
  publisher = {Association for Computing Machinery},
  address   = {New York, NY, USA},
  pages     = {10--19},
  year      = {2003},
  doi       = {10.1145/780542.780545},
  url       = {https://doi.org/10.1145/780542.780545}
}

@article{McMullen1970,
  author  = {McMullen, P.},
  title   = {The maximum numbers of faces of a convex polytope},
  journal = {Mathematika},
  volume  = {17},
  number  = {2},
  pages   = {179--184},
  year    = {1970},
  doi     = {10.1112/S0025579300002850},
  url     = {https://doi.org/10.1112/S0025579300002850}
}

@article{GavinskyIto2013,
  author  = {Gavinsky, Dmitry and Ito, Tsuyoshi},
  title   = {Quantum fingerprints that keep secrets},
  journal = {Quantum Inf. Comput.},
  volume  = {13},
  number  = {7\&8},
  pages   = {583--606},
  year    = {2013},
  doi     = {10.26421/QIC13.7-8-3},
  url     = {https://doi.org/10.26421/QIC13.7-8-3}
}

@article{Gharibian2010,
  author  = {Gharibian, Sevag},
  title   = {Strong {NP}-hardness of the quantum separability problem},
  journal = {Quantum Inf. Comput.},
  volume  = {10},
  number  = {3\&4},
  pages   = {343--360},
  year    = {2010},
  doi     = {10.26421/QIC10.3-4-11},
  url     = {https://doi.org/10.26421/QIC10.3-4-11}
}

@article{ZhangReassessing2026,
  author  = {Zhang, Yujie and Schmid, David and Y{\=\i}ng, Y{\`i}l{\`e} and Spekkens, Robert W.},
  title   = {Reassessing the Boundary between Classical and Nonclassical for Individual Quantum Processes},
  journal = {Phys. Rev. X},
  volume  = {16},
  number  = {2},
  pages   = {021050},
  year    = {2026},
  doi     = {10.1103/vqfz-wzjg},
  url     = {https://doi.org/10.1103/vqfz-wzjg}
}

@article{Szarek1998,
  author  = {Szarek, Stanis{\l}aw},
  title   = {Metric Entropy of Homogeneous Spaces},
  journal = {Banach Center Publ.},
  volume  = {43},
  pages   = {395--410},
  year    = {1998},
  doi     = {10.4064/-43-1-395-410},
  url     = {https://doi.org/10.4064/-43-1-395-410}
}

@article{Busch2003,
  title = {Quantum States and Generalized Observables: A Simple Proof of Gleason's Theorem},
  author = {Busch, P.},
  journal = {Phys. Rev. Lett.},
  volume = {91},
  issue = {12},
  pages = {120403},
  numpages = {4},
  year = {2003},
  month = {Sep},
  publisher = {American Physical Society},
  doi = {10.1103/PhysRevLett.91.120403},
  url = {https://link.aps.org/doi/10.1103/PhysRevLett.91.120403}
}

@book{AubrunSzarek2017,
  author    = {Aubrun, Guillaume and Szarek, Stanis{\l}aw J.},
  title     = {Alice and Bob Meet Banach: The Interface of Asymptotic Geometric Analysis and Quantum Information Theory},
  series    = {Mathematical Surveys and Monographs},
  volume    = {223},
  publisher = {American Mathematical Society},
  year      = {2017},
  isbn      = {978-1-4704-3468-7},
  url       = {https://bookstore.ams.org/surv-223}
}

@article{Zhang2026quantifiers,
  doi = {10.22331/q-2026-07-30-2180},
  url = {https://doi.org/10.22331/q-2026-07-30-2180},
  title = {Quantifiers and witnesses for the nonclassicality of measurements and of states},
  author = {Zhang, Yujie and Yīng, Y{\`{i}}l{\`{e}} and Schmid, David},
  journal = {{Quantum}},
  issn = {2521-327X},
  publisher = {{Verein zur F{\"{o}}rderung des Open Access Publizierens in den Quantenwissenschaften}},
  volume = {10},
  pages = {2180},
  month = jul,
  year = {2026}
}

@article{MontanaroDeWolf2016,
  author  = {Montanaro, Ashley and de Wolf, Ronald},
  title   = {A Survey of Quantum Property Testing},
  journal = {Theory Comput. Grad. Surv.},
  volume  = {7},
  pages   = {1--81},
  year    = {2016},
  doi     = {10.4086/toc.gs.2016.007},
  url     = {https://doi.org/10.4086/toc.gs.2016.007}
}

@article{GuhneToth2009,
  author  = {G{\"u}hne, Otfried and T{\'o}th, G{\'e}za},
  title   = {Entanglement detection},
  journal = {Phys. Rep.},
  volume  = {474},
  number  = {1--6},
  pages   = {1--75},
  year    = {2009},
  doi     = {10.1016/j.physrep.2009.02.004},
  url     = {https://doi.org/10.1016/j.physrep.2009.02.004}
}
 
\clearpage
\onecolumngrid
\setcounter{page}{1}
\renewcommand{\thepage}{S\arabic{page}}
\setcounter{equation}{0}
\renewcommand{\theequation}{S\arabic{equation}}
\renewcommand{\theHequation}{S\arabic{equation}}
\setcounter{section}{0}
\renewcommand{\thesection}{S\arabic{section}}
\renewcommand{\theHsection}{S\arabic{section}}
\setcounter{lemma}{0}
\renewcommand{\thelemma}{S\arabic{lemma}}
\renewcommand{\theHlemma}{S\arabic{lemma}}
\begin{center}
{\large\bf Supplemental Material}
\end{center}
 
We use the notation of the main text. The Haar measure $\mu$ on pure states is normalized to one, $P_u=\proj{u}$, and logarithms are natural except in $E_{\max}$.
 
\section{Finite tests and entanglement resolution}
\label{sec:finite-characterization-sm}
 
\subsection{Proof of Lemma~\ref{lem:finite-test}}
\label{sec:proof-finite-test-sm}
 
Suppose first that the statistics of $\rho^{AB}$ admit a decomposition of the form of Eq.~\eqref{eq:finite-nc-model}. Since $p(a|x,\lambda_A)$ preserves every identity in $\mc O_{\rm all}(\msf N)$, the assignment $N^A_{a|x}\mapsto p(a|x,\lambda_A)$ extends to a well-defined real-linear functional on the real span of Alice's effects~\cite{SchmidSelbyPuseySpekkens2024,ZhangEntCert2025}. If $\msf N$ is informationally complete, this span is $\Herm(\mc H^A)$, and the functional has a unique Hermitian representative $F_{\lambda_A}$, i.e., $p(a|x,\lambda_A)=\tr(F_{\lambda_A}N^A_{a|x})$. Normalization, together with $\sum_aN^A_{a|x}=\mbb 1^A$, gives $\tr F_{\lambda_A}=1$, and thus $F_{\lambda_A}\in\mc K_A$. Similarly, $p(b|y,\lambda_B)=\tr(G_{\lambda_B}M^B_{b|y})$ with $G_{\lambda_B}\in\mc K_B$. Substituting these representations into Eq.~\eqref{eq:finite-nc-model} gives
\begin{equation}
\tr\left[(N_{a|x}^A\otimes M_{b|y}^B)\left(\rho^{AB}-\sum_{\lambda_A,\lambda_B}p(\lambda_A,\lambda_B)F_{\lambda_A}\otimes G_{\lambda_B}\right)\right]=0
\label{eq:response-product-sm}
\end{equation}
for all $a,b,x,y$. Since the product effects are informationally complete on $AB$, it follows that $\rho^{AB}=\sum_{\lambda_A\lambda_B}p(\lambda_A\lambda_B)F_{\lambda_A}\otimes G_{\lambda_B}\in\mc C_{\msf N,\msf M}$.
 
Conversely, suppose that $\rho^{AB}=\rho^{AB}=\sum_{\lambda_A\lambda_B}p(\lambda_A\lambda_B)F_{\lambda_A}\otimes G_{\lambda_B}$ with $F_{\lambda_A}\in\mc K_A$ and $G_{\lambda_B}\in\mc K_B$. Taking $p(a|x,\lambda_A):=\tr(F_{\lambda_A}N^A_{a|x})$ and $p(b|y,\lambda_B):=\tr(G_{\lambda_B}M^B_{b|y})$, we obtain response functions that are nonnegative and normalized, preserve all local identities by linearity, and reproduce Eq.~\eqref{eq:finite-statistics}.
 
 
\subsection{Proof of Theorem~\ref{thm:resolution-covering}}
\label{sec:proof-resolution-covering-sm}
 
\paragraph{Upper bound.}
Let $F\in\mc K_A$, and let $P_u$ project onto an eigenvector corresponding to $\lambda_{\min}(F)$. By the definition of $\epsilon_A$, there exists an Alice effect $Q$ such that
\begin{equation}
1-\frac{\tr(P_uQ)}{\tr Q}\leq\epsilon_A.
\end{equation}
If $F$ is not positive, then $\lambda_{\min}(F)<0<\lambda_{\max}(F)$. Since $\tr F=1$ gives $\lambda_{\max}(F)+(d-1)\lambda_{\min}(F)\leq1$, we obtain
\begin{equation}
0\leq\frac{\tr(FQ)}{\tr Q}\leq(1-\epsilon_A)\lambda_{\min}(F)+\epsilon_A\lambda_{\max}(F)\leq\epsilon_A+(1-d\epsilon_A)\lambda_{\min}(F).
\label{eq:spectral-repair-sm}
\end{equation}
Therefore, whenever $d\epsilon_A<1$,
\begin{equation}
\lambda_{\min}(F)\geq-\frac{\epsilon_A}{1-d\epsilon_A}\quad\Longrightarrow\quad \Delta_{d\epsilon_A}(F)=(1-d\epsilon_A)F+\epsilon_A\mbb 1^A\geq\mbb 0^A,
\label{eq:min-eigenvalue-bound-sm}
\end{equation}
where $\Delta_t(X):=(1-t)X+t\tr(X)\mbb 1/d$ is the depolarizing channel, so that $\Delta_{d\epsilon_A}(F)$ is a density operator. Analogously, $\Delta_{d\epsilon_B}(G)$ is a density operator for every $G\in\mc K_B$.

Now let $\rho^{AB}\in\mc C_{\msf N,\msf M}$. Applying $\Delta_{d\epsilon_A}\otimes\Delta_{d\epsilon_B}$ to a pseudo-product decomposition of $\rho$ maps every local pseudo-state to a density operator, and therefore produces a separable state $\sigma$. Expanding the depolarizing maps gives
\begin{equation}
\sigma^{AB}=(\Delta_{d\epsilon_A}\otimes\Delta_{d\epsilon_B})(\rho^{AB})=(1-d\epsilon_A)(1-d\epsilon_B)\rho^{AB}+\epsilon_A(1-d\epsilon_B)\mbb 1^A\otimes\rho^B+\epsilon_B(1-d\epsilon_A)\rho^A\otimes\mbb 1^B+\epsilon_A\epsilon_B\mbb 1^{AB},
\end{equation}
where $\rho^A=\tr[\rho^{AB}]$, $\rho_B=\tr_A[\rho^{AB}]$ and hence the last three terms are separable states.  By the triangle inequality for the trace norm,
\begin{equation}
E_{\rm tr}(\rho^{AB})\leq\frac12\norm{\rho^{AB}-\sigma^{AB}}_1\leq1-(1-d\epsilon_A)(1-d\epsilon_B)\leq2d\epsilon.
\label{eq:resolution-upper-sm}
\end{equation}
Taking the maximum over $\rho^{AB}\in\mc C_{\msf N,\msf M}$ proves the upper bound with $C_d=2d$.
 
\paragraph{The Werner family.}
We write the states of Eq.~\eqref{eq:werner-hard-main} as
\begin{equation}
\rho^{AB}_\gamma=\alpha\mbb 1^{AB}-\beta\mbb S^{AB},\qquad \alpha:=\frac{d+2\gamma}{d(d^2-1)},\qquad \beta:=\frac{1+2d\gamma}{d(d^2-1)}.
\label{eq:werner-hard-sm}
\end{equation}
Their eigenvalues are $(1-2\gamma)/[d(d+1)]$ on the symmetric subspace and $(1+2\gamma)/[d(d-1)]$ on the antisymmetric subspace, so that $\rho^{AB}_\gamma\geq0$ for $-1/2\leq\gamma\leq1/2$, and $\tr(\mbb S\rho^{AB}_\gamma)=-2\gamma$. 

Werner states with $\gamma\le 0$ are separable~\cite{Werner1989}, so that $E_{\rm tr}(\rho^{AB}_\gamma)=0$ for $\gamma\leq0$. For $0\leq\gamma\leq1/2$, the eigenvalues give $D_{\rm tr}(\rho^{AB}_\gamma,\rho^{AB}_0)=\gamma$, hence $E_{\rm tr}(\rho^{AB}_\gamma)\leq\gamma$. Conversely, $\tr(\mbb S^{AB}\sigma^{AB})\geq0$ for every separable $\sigma^{AB}$ and $\norm{\mbb S^{AB}}_\infty=1$ give $D_{\rm tr}(\rho^{AB}_\gamma,\sigma^{AB})\geq\frac12|\tr[\mbb S(\rho^{AB}_\gamma-\sigma^{AB})]|\geq\gamma$. Thus
\begin{equation}
E_{\rm tr}(\rho^{AB}_\gamma)=\gamma\qquad(0\leq\gamma\leq1/2).
\label{eq:werner-etr-sm}
\end{equation}

\paragraph{Lower bound}

Without loss of generality, $\epsilon_A=\epsilon$. Choose a worst-covered pure state $P_u=\ket u\bra u$, so that $\tr(P_uQ)/\tr Q\leq1-\epsilon$ for every nonzero Alice effect $Q$, and set
\begin{equation}
q_*:=\frac{d-1}{4d(d+1)},\qquad t:=\frac{\epsilon}{d-1},\qquad \gamma:=\frac{q_*t}{4}=\frac{\epsilon}{16d(d+1)}.
\label{eq:werner-resolution-parameters-sm}
\end{equation}
The trace-one operator $F_{u,t}:=-tP_u+(1+t)(\mbb 1-P_u)/(d-1)$ belongs to $\mc K_A$, since
\begin{equation}
\frac{\tr(F_{u,t}Q)}{\tr Q}\geq-t+\frac{1+dt}{d-1}\epsilon=\frac{d\epsilon^2}{(d-1)^2}\geq0.
\end{equation}
To prove $\rho^{AB}\in \mc C_{\msf N,\msf M}$, then it suffices to prove that $R:=\rho^{AB}_\gamma-q_*F_{u,t}\otimes P_u$ is separable, since $F_{u,t}\otimes P_u\in \mc C_{\msf N,\msf M}$.

Choose a common local orthonormal basis with $\ket u$, and put $\delta:=d\beta-\alpha=2\gamma/d$. For $a,b\geq0$, define
\begin{equation}
B_{ij}(a,b):=a\proj{ii}+b\proj{jj}+\beta(\ket{ij}-\ket{ji})(\bra{ij}-\bra{ji}).
\label{eq:separable-block-sm}
\end{equation}
This operator is separable by the two-qubit PPT criterion whenever $ab\geq\beta^2$. Our parameters give $\delta\leq\beta/2$, so both $B_{ij}(\beta,\beta)$ and $B_{ij}(\beta+2\delta,\beta-\delta)$ are separable, as $(\beta+2\delta)(\beta-\delta)-\beta^2=\delta(\beta-2\delta)\geq0$. Direct expansion gives
\begin{equation}
R=\sum_{2\leq i<j\leq d}B_{ij}(\beta,\beta)+\sum_{j=2}^dB_{1j}(\beta+2\delta,\beta-\delta)+D,
\end{equation}
where $D=\delta\proj{11}+\left(\alpha-\beta-\frac{q_*(1+t)}{d-1}\right)\sum_{i=2}^d\proj{i1}+(\alpha-\beta)\sum_{\substack{i\neq j\\j\geq2}}\proj{ij}$ for small $\epsilon$. All coefficients are nonnegative, since $\alpha-\beta-q_*(1+t)/(d-1)=(3-t-8\gamma)/[4d(d+1)]\geq0$. Hence $D$ and $R$ are separable.

Thus $\rho^{AB}_\gamma\in\mc C_{\msf N,\msf M}$, since $\rho^{AB}_\gamma$ remains uncertified for every measurement configuration with covering infidelity $\epsilon$. Consequently,
\begin{equation}
\gamma^{\rm tr}_{\msf N,\msf M}\geq E_{\rm tr}(\rho^{AB}_\gamma)=\frac{\epsilon}{16d(d+1)}.
\label{eq:werner-resolution-lower-sm}
\end{equation}
Together with the upper bound, this proves Theorem~\ref{thm:resolution-covering} with
\begin{equation}
c_d=\frac{1}{16d(d+1)},\qquad C_d=2d.
\label{eq:resolution-explicit-constants-sm}
\end{equation}

\subsection{Other entanglement measures}
\label{sec:measure-comparisons-sm}

Generalized robustness is $E_{\rm gr}(\rho^{AB})=\min\{s\geq0:\rho^{AB}+s\tau^{AB}=(1+s)\sigma^{AB},\ \tau^{AB}\in\mc D_{AB},\ \sigma^{AB}\in\Sep\}$~\cite{Steiner2003}. Any feasible decomposition gives $E_{\rm tr}(\rho^{AB})\leq s/(1+s)$, hence $E_{\rm tr}(\rho)\leq E_{\rm gr}(\rho^{AB})/[1+E_{\rm gr}(\rho^{AB})]$. Conversely, its witness dual~\cite{Brandao2005} is
\begin{equation}
E_{\rm gr}(\rho^{AB})=\max\{-\tr(W^{AB}\rho^{AB}):W^{AB}\leq\mbb 1^{AB},\ \tr(W^{AB}\sigma^{AB})\geq0\ \forall\sigma^{AB}\in\Sep\}.
\label{eq:gr-dual-sm}
\end{equation}
For feasible $W^{AB}$, set $X^{AB}=\mbb 1^{AB}-W^{AB}\geq0$. Every product expectation of $^{AB}X$ is at most one, while a top eigenvector has squared product overlap at least $1/d$, so $\norm X^{AB}_\infty\leq d$. If $\sigma^{AB}_*$ is closest separable, then
\begin{equation}
-\tr(W^{AB}\rho^{AB})\leq-\tr[W^{AB}(\rho^{AB}-\sigma^{AB}_*)]=\tr[X^{AB}(\rho^{AB}-\sigma^{AB}_*)]\leq dE_{\rm tr}(\rho^{AB}).
\end{equation}
Therefore
\begin{equation}
\frac{1}{d}E_{\rm gr}(\rho^{AB})\leq E_{\rm tr}(\rho^{AB})\leq\frac{E_{\rm gr}(\rho^{AB})}{1+E_{\rm gr}(\rho^{AB})}.
\label{eq:tr-gr-comparison-sm}
\end{equation}
In particular, $E_{\rm tr}(\rho^{AB})\leq E_{\rm gr}(\rho^{AB})\leq dE_{\rm tr}(\rho^{AB})$, so taking the maximum over $\rho\in\mc C_{\msf N,\msf M}$ and using Theorem~\ref{thm:resolution-covering} gives $\gamma_{\msf N,\msf M}^{\rm gr}=\Theta_d(\epsilon)$. Since $E_{\max}(\rho^{AB})=\log_2[1+E_{\rm gr}(\rho^{AB})]$, the expansion $\log_2(1+x)=x/\ln2+O(x^2)$ also gives $\gamma_{\msf N,\msf M}^{\max}=\Theta_d(\epsilon)$.

\section{Effect complexity}
\label{sec:optimal-effect-complexity-sm}
 
Throughout this section, $d\geq2$ is fixed. For a nonzero effect $Q$, let
\begin{equation}
\mc U_\epsilon(Q):=\{P_u:\tr(P_uQ)/\tr Q\geq1-\epsilon\}
\label{eq:effect-coverage-set-sm}
\end{equation}
denote the set of pure states that $Q$ covers within infidelity $\epsilon$. For a rank-one effect $Q=wP_v$, this is the closed trace-distance ball of radius $\sqrt\epsilon$ around $P_v$. For a Haar-random pure state $P_u$, the overlap $x=\tr(P_uP_v)$ has density $(d-1)(1-x)^{d-2}$ on $[0,1]$~\cite{ZyczkowskiSommers2005}. Hence, for $0\leq\epsilon\leq1$,
\begin{equation}
\mc U_\epsilon(wP_v)=\{P_u:D_{\rm tr}(P_u,P_v)\leq\sqrt\epsilon\},\qquad \mu[\mc U_\epsilon(P_v)]=\int_{1-\epsilon}^1(d-1)(1-x)^{d-2}\,dx=\epsilon^{d-1},
\label{eq:pure-state-ball-volume-sm}
\end{equation}
in agreement with the volume of trace-distance balls of pure states given in Ref.~\cite{OszmaniecEtAl2024}.
 
For $0<\epsilon\leq1$, we choose projectors $\{P_{v_j}\}_{j=1}^L$ with pairwise trace distances greater than $\sqrt\epsilon$, until no further projector can be added. The balls of radius $\sqrt\epsilon$ around them cover all pure states, since otherwise an uncovered state could be added. The balls of radius $\sqrt\epsilon/2$, i.e., the sets $\mc U_{\epsilon/4}(P_{v_j})$, are pairwise disjoint by the triangle inequality. Using Eq.~\eqref{eq:pure-state-ball-volume-sm}, the covering property gives
\begin{equation}
1\leq\sum_{j=1}^L\mu[\mc U_\epsilon(P_{v_j})]=L\epsilon^{d-1},
\label{eq:covering-volume-bound-sm}
\end{equation}
while disjointness gives
\begin{equation}
1\geq\sum_{j=1}^L\mu[\mc U_{\epsilon/4}(P_{v_j})]=L\left(\frac{\epsilon}{4}\right)^{d-1}.
\label{eq:disjoint-volume-bound-sm}
\end{equation}
Consequently,
\begin{equation}
\left(\frac{1}{\epsilon}\right)^{d-1}\leq L\leq\left(\frac{4}{\epsilon}\right)^{d-1}.
\label{eq:pure-state-packing-covering-sm}
\end{equation}
This is the standard volumetric packing and covering argument; see, e.g., Ref.~\cite{AubrunSzarek2017}.
 
\subsection{Proof of Lemma~\ref{lem:effect-covering}}
\label{sec:proof-effect-covering-sm}
 
\paragraph{Lower bound on $\epsilon_{\rm opt}(n)$.}
We first show that, for $\epsilon<1/2$, a single effect covers no more pure states than a rank-one effect. Fix a nonzero effect $Q$ with $\mc U_\epsilon(Q)\neq\varnothing$; otherwise there is nothing to prove. Let $1-a$ be the largest eigenvalue of $Q/\tr Q$, with eigenvector $\ket v$. Since $\tr(P_uQ)/\tr Q\leq1-a$ for every $P_u$, nonemptiness implies $a\leq\epsilon$. For $P_u\in\mc U_\epsilon(Q)$, write $x=\tr(P_uP_v)$. The remaining eigenvalues of $Q/\tr Q$ are nonnegative, sum to $a$, and belong to eigenvectors orthogonal to $\ket v$, so that
\begin{equation}
1-\epsilon\leq\frac{\tr(P_uQ)}{\tr Q}\leq(1-a)x+a(1-x).
\end{equation}
Using $0\leq a\leq\epsilon<1/2$, we obtain
\begin{equation}
1-x\leq\frac{\epsilon-a}{1-2a}\leq\epsilon,\qquad\text{i.e.,}\qquad \mc U_\epsilon(Q)\subseteq\mc U_\epsilon(P_v).
\label{eq:rank-one-domination-sm}
\end{equation}
Equation~\eqref{eq:pure-state-ball-volume-sm} therefore gives $\mu[\mc U_\epsilon(Q)]\leq\epsilon^{d-1}$ for every nonzero effect.
 
Now let $\msf N$ have at most $n$ cone-inequivalent effects and covering infidelity $\epsilon<1/2$, and select one nonzero effect $Q_i$ from each cone class. Since cone-equivalent effects cover the same pure states, the sets $\mc U_\epsilon(Q_i)$ cover all pure states, and
\begin{equation}
1=\mu \left[\bigcup_i\mc U_\epsilon(Q_i)\right]\leq\sum_i\mu[\mc U_\epsilon(Q_i)]\leq n\epsilon^{d-1},
\end{equation}
i.e., $\epsilon\geq n^{-1/(d-1)}$. If instead $\epsilon\geq1/2$, then $\epsilon\geq\frac12n^{-1/(d-1)}$ holds trivially. Hence every multi-measurement with at most $n$ cone classes satisfies
\begin{equation}
\epsilon_A(\msf N)\geq\frac12n^{-1/(d-1)},
\label{eq:effect-lower-explicit-sm}
\end{equation}
and $\epsilon_{\rm opt}(n)=\Omega_d(n^{-1/(d-1)})$. This argument uses only the positivity of the $Q_i$, and not that they form measurements.
 
\paragraph{Upper bound on $\epsilon_{\rm opt}(n)$.}
Given $0<\epsilon\leq1$, take a covering family of at most $(4/\epsilon)^{d-1}$ projectors as in Eq.~\eqref{eq:pure-state-packing-covering-sm}. We complete each defining vector to an orthonormal basis and include the corresponding rank-one projective measurement. The resulting multi-measurement contains all covering projectors, so that its covering infidelity is at most $\epsilon$, and it has at most $d(4/\epsilon)^{d-1}$ effects. For $n\geq d4^{d-1}$, choosing $\epsilon=4(d/n)^{1/(d-1)}\leq1$ makes this number at most $n$, which proves $\epsilon_{\rm opt}(n)\leq4d^{1/(d-1)}n^{-1/(d-1)}$.

\subsection{Proof of Lemma~\ref{lem:haar-covering}}
\label{sec:proof-haar-covering-sm}
 
We first prove the statement for $n$ independent Haar-random pure states $P_1,\ldots,P_n$ and then extend it to projective measurements, which is the case stated in Lemma~\ref{lem:haar-covering}. Since $1-\tr(P_uQ)/\tr Q=1-\tr(P_uP_i)=D_{\rm tr}(P_u,P_i)^2$ for every effect $Q\propto P_i$, the covering infidelity depends only on the directions $P_i$, and we also denote it by $\epsilon_{\rm Haar}(n)$ in this case. For each $\epsilon$, we fix the auxiliary pure states $\{P_{v_j}\}_{j=1}^L$ of Eq.~\eqref{eq:pure-state-packing-covering-sm} before sampling. Their balls $\mc U_\epsilon(P_{v_j})$ cover all pure states, while the smaller balls $\mc U_{\epsilon/4}(P_{v_j})$ are pairwise disjoint. We call a ball empty if it contains none of $P_1,\ldots,P_n$. For any fixed $P_v$, independence and Eq.~\eqref{eq:pure-state-ball-volume-sm} give
\begin{equation}
\Pr[P_i\notin\mc U_\epsilon(P_v)\ \forall i]=(1-\epsilon^{d-1})^n\leq e^{-n\epsilon^{d-1}}.
\label{eq:block-empty-sm}
\end{equation}
 
\paragraph{Lower bound on $\epsilon_{\rm Haar}(n)$.}
Set $\epsilon=[\log n/(2n)]^{1/(d-1)}$. We show that, with probability tending to one, some auxiliary state $P_{v_j}$ has infidelity larger than $\epsilon/4$ with every sampled direction, i.e., that some smaller ball $\mc U_{\epsilon/4}(P_{v_j})$ is empty. By Eq.~\eqref{eq:covering-volume-bound-sm},
\begin{equation}
L\geq\epsilon^{-(d-1)}=\frac{2n}{\log n}.
\end{equation}
Let $I_j=1$ if $\mc U_{\epsilon/4}(P_{v_j})$ is empty and $I_j=0$ otherwise, and let $Z:=\sum_{j=1}^L I_j$ count the empty smaller balls. Each has measure $(\epsilon/4)^{d-1}$, so that
\begin{equation}
\mathbb E(Z)=L[1-(\epsilon/4)^{d-1}]^n\geq L(1-\epsilon^{d-1})^n=\Omega \left(\frac{\sqrt n}{\log n}\right)\longrightarrow\infty,
\end{equation}
where we used $(1-\epsilon^{d-1})^n=[1-\log n/(2n)]^n=n^{-1/2}(1+o(1))$. To bound the variance of $Z$, note that, since two smaller balls are disjoint, the probability that all sampled directions miss both of them is
\begin{equation}
\mathbb E(I_jI_\ell)=[1-2(\epsilon/4)^{d-1}]^n\leq[1-(\epsilon/4)^{d-1}]^{2n}=\mathbb E(I_j)\mathbb E(I_\ell)\qquad(j\ne\ell).
\label{eq:haar-second-moment-sm}
\end{equation}
Hence $\text{Var}(Z)\leq\sum_j\text{Var}(I_j)\leq\mathbb E(Z)$, and Chebyshev's inequality gives
\begin{equation}
\Pr[Z=0]\leq\frac{\text{Var}(Z)}{[\mathbb E(Z)]^2}\leq\frac1{\mathbb E(Z)}=O \left(\frac{\log n}{\sqrt n}\right)\longrightarrow0.
\end{equation}
When $Z>0$, some $P_{v_j}$ has infidelity greater than $\epsilon/4$ with every $P_i$, so that, with probability tending to one,
\begin{equation}
\epsilon_{\rm Haar}(n)=\max_{P_u}\min_{1\leq i\leq n}[1-\tr(P_uP_i)]\geq\min_{1\leq i\leq n}[1-\tr(P_{v_j}P_i)]>\frac{\epsilon}{4}.
\end{equation}
 
\paragraph{Upper bound on $\epsilon_{\rm Haar}(n)$.}
For $0<\epsilon\leq1$, we now use the larger balls $\mc U_\epsilon(P_{v_j})$, which cover all pure states and number at most $L\leq(4/\epsilon)^{d-1}$ by Eq.~\eqref{eq:disjoint-volume-bound-sm}. Suppose that every larger ball contains at least one sampled direction. For any pure state $P_u$, choose $P_{v_j}$ whose ball contains $P_u$, and a sampled direction $P_i$ in that ball. The triangle inequality gives
\begin{equation}
D_{\rm tr}(P_u,P_i)\leq D_{\rm tr}(P_u,P_{v_j})+D_{\rm tr}(P_{v_j},P_i)\leq2\sqrt\epsilon,
\end{equation}
so that every pure state has infidelity at most $4\epsilon$ with some sampled direction, i.e., $\epsilon_{\rm Haar}(n)\leq4\epsilon$. Consequently, $\epsilon_{\rm Haar}(n)>4\epsilon$ requires at least one larger ball to be empty, and Eq.~\eqref{eq:block-empty-sm} with a union bound gives
\begin{equation}
\Pr[\epsilon_{\rm Haar}(n)>4\epsilon]\leq(4/\epsilon)^{d-1}e^{-n\epsilon^{d-1}}.
\end{equation}
Taking $\epsilon=(\log n/n)^{1/(d-1)}$ gives
\begin{equation}
\Pr \left[\epsilon_{\rm Haar}(n)>4\left(\frac{\log n}{n}\right)^{1/(d-1)}\right]\leq\frac{4^{d-1}}{\log n}\longrightarrow0.
\label{eq:haar-upper-sm}
\end{equation}
Combining the two bounds,
\begin{equation}
\Pr \left[\frac14\left(\frac{\log n}{2n}\right)^{1/(d-1)}<\epsilon_{\rm Haar}(n)\leq4\left(\frac{\log n}{n}\right)^{1/(d-1)}\right]\longrightarrow1\qquad(n\to\infty).
\label{eq:haar-two-sided-sm}
\end{equation}
 
\paragraph{Extension to Haar-random PVMs.}
Let $m=n/d$ independent Haar-random PVMs be sampled. For $\epsilon<1/2$, orthogonality allows at most one projector from each PVM in $\mc U_\epsilon(P_v)$. Since each projector is Haar distributed, one PVM hits this ball with probability $d\epsilon^{d-1}$. Its empty probability is therefore $(1-d\epsilon^{d-1})^m\leq e^{-n\epsilon^{d-1}}$, so the upper-bound proof applies unchanged.

For the lower bound, it suffices to find an uncovered direction near one fixed pure state. Restrict the auxiliary states in the preceding construction to a fixed ball $U$ of radius $1/4$. Their larger balls cover $U$, so $\mu(U)\leq L\epsilon^{d-1}$ still gives $L=\Omega_d(\epsilon^{-(d-1)})$. The smaller balls are disjoint, and any two states in their union have distance at most $1/2+\sqrt\epsilon<1$ for sufficiently small $\epsilon$. Thus their union contains no orthogonal pair, preventing one PVM from hitting two different smaller balls, so the lower-bound proof applies analogously. 
\label{sec:known-identities-sm}

\section{Sample complexity}
\label{sec:known-identities-sm}
 
We fix an entanglement resolution $\gamma>0$ and an error probability $0<\delta\leq1/4$. The exact operational identities $\mc O_{\rm all}(\msf N)$ and $\mc O_{\rm all}(\msf M)$ are assumed to be established independently of the target data, for the same stationary devices used throughout the protocol, and they determine $\mc{NC}_{\rm aattt}(\mc B)$. We count only copies of the target state.
 
\begin{protocol}{Certification with independently established identities}
\label{prot:fixed-identities}
\begin{protocolsteps}
\item From the supplied identities, construct the polytopes $\mbb P_{\msf N}$, $\mbb P_{\msf M}$, and $\mc{NC}_{\rm aattt}(\mc B)$.
\item Using only the identities, select $d^2$ linearly independent effects on each side, $\{N_k\}_{k=1}^{d^2}\subseteq\msf N$ and $\{M_l\}_{l=1}^{d^2}\subseteq\msf M$, whose coordinate matrix has maximal absolute determinant. Let $\mc J$ index the $d^4$ product events $j=(k,l)$, and let $\mc S$ be the set of setting pairs needed to observe them.
\item Choose the number $K$ of repetitions per setting pair according to Eq.~\eqref{eq:fixed-identity-allocation-sm}. For every $(x,y)\in\mc S$, measure $K$ new copies of the target state and record all outcomes.
\item For each $j\in\mc J$, let $f_j$ be the fraction of the $K$ runs at its setting pair that produced its outcome pair, and compute the confidence radius $r$ of Eq.~\eqref{eq:raw-target-region-sm}.
\item Return $\mathsf{certify}$ if no $q\in\mc{NC}_{\rm aattt}(\mc B)$ satisfies $|q_j-f_j|\leq r$ for all $j\in\mc J$; otherwise, return $\mathsf{inconclusive}$.
\end{protocolsteps}
\end{protocol}
 
We make two remarks about step~2. First, the identities determine the effect vectors up to an invertible linear map on their span, and such a map multiplies the determinants of all $d^2$-element subsets by the same factor. The maximizing subset can therefore be found from the identities alone, in any coordinates consistent with them. Second, any subset whose absolute determinant is at least a fixed fraction of the maximum can be used in the analysis below, with modified constants.
 
\paragraph{Soundness.}
The total number of target copies is $N=|\mc S|K$, since events belonging to the same setting pair are estimated from the same $K$ copies. For each $j\in\mc J$, $f_j$ is an average of $K$ independent Bernoulli variables with mean $p_j$, and Hoeffding's inequality~\cite{Hoeffding1963} gives $\Pr[|f_j-p_j|>r]\leq2e^{-2Kr^2}$. A union bound over the $d^4$ events yields
\begin{equation}
\Pr \left[\max_{j\in\mc J}|f_j-p_j|>r\right]\leq2d^4e^{-2Kr^2}.
\end{equation}
Setting the right-hand side equal to $\delta$ gives the confidence radius and region
\begin{equation}
r:=\sqrt{\frac{\log(2d^4/\delta)}{2K}},\qquad \mc R_N:=\{q:|q_j-f_j|\leq r\ \forall j\in\mc J\}.
\label{eq:raw-target-region-sm}
\end{equation}
Thus $p\in\mc R_N$ with probability at least $1-\delta$. If $p\in\mc{NC}_{\rm aattt}(\mc B)$, then on this event $q=p$ is a point of $\mc{NC}_{\rm aattt}(\mc B)$ within the confidence region, and the protocol does not certify. Hence false certification has probability at most $\delta$. Note that this argument uses only the supplied identities.
 
The decision in step~5 is a linear feasibility problem~\cite{SchmidSpekkensWolfe2018}. Writing $\xi^A_\ell$ and $\xi^B_{\ell'}$ for the vertices of $\mbb P_{\msf N}$ and $\mbb P_{\msf M}$, it asks whether there exist weights $w_{\ell\ell'}\geq0$ with $\sum_{\ell,\ell'}w_{\ell\ell'}=1$ such that
\begin{equation}
f_j-r\leq\sum_{\ell,\ell'}w_{\ell\ell'}(\xi^A_\ell\otimes\xi^B_{\ell'})_j\leq f_j+r\qquad(j\in\mc J).
\label{eq:fixed-identity-data-lp-sm}
\end{equation}
All implemented effects constrain these polytopes, including those that are not sampled on the target. In a numerical implementation, $\mathsf{certify}$ should be returned only if infeasibility is verified.
 
\subsection{Upper bound on the target-copy count}
\label{sec:known-identity-proof-sm}

For a fixed sufficiently small resolution $\gamma>0$, choose the rank-one projective designs of Lemma~\ref{lem:effect-covering} with covering infidelity $\epsilon\leq\gamma/(16d)$. These use $O_d(\gamma^{-(d-1)})$ effects.

Let $\mc D=\Delta_{d\epsilon}\otimes\Delta_{d\epsilon}$ be the local depolarizing map from the earlier repair argument. For every $X\in\mc T_{\msf N,\msf M}$, including nonpositive $X$, we have $\mc D(X)\in\Sep$. The earlier disturbance bound gives $\norm{\rho-\mc D(\rho)}_1\leq4d\epsilon$. Hence, whenever $E_{\rm tr}(\rho)\geq\gamma$, trace-norm contractivity and the triangle inequality give
\begin{align}
\norm{\rho-X}_1\geq\norm{\mc D(\rho)-\mc D(X)}_1\geq\norm{\rho-\mc D(X)}_1-\norm{\rho-\mc D(\rho)}_1\geq2\gamma-4d\epsilon\geq\gamma.
\label{eq:target-full-nc-trace-gap-sm}
\end{align}

Write $\{N_\alpha\}_{\alpha=1}^{d^2}$ and $\{M_\beta\}_{\beta=1}^{d^2}$ for the selected local effect bases. For each retained event $j=(\alpha,\beta)\in\mc J$, denote its product effect by $E_j:=N_\alpha\otimes M_\beta$. Informational completeness gives the norm-equivalence bound
\begin{align}
\max_{j\in\mc J}\left|\tr(E_jH)\right|\geq\nu_d\norm{H}_1\qquad\text{for every Hermitian }H.
\label{eq:fixed-identity-norming-sm}
\end{align}
The determinant-maximizing selection ensures that $\nu_d>0$ can be chosen uniformly over these covering designs, depending only on $d$.

The target probabilities are $p_j=\tr(E_j\rho)$. Every noncontextual candidate $q$ is represented by some $X\in\mc T_{\msf N,\msf M}$, with $q_j=\tr(E_jX)$. Combining Eqs.~\eqref{eq:target-full-nc-trace-gap-sm} and~\eqref{eq:fixed-identity-norming-sm} therefore gives
\begin{align}
\max_{j\in\mc J}|q_j-p_j|=\max_{j\in\mc J}\left|\tr[E_j(X-\rho)]\right|\geq\nu_d\norm{X-\rho}_1\geq\nu_d\gamma.
\label{eq:fixed-identity-probability-gap-sm}
\end{align}
Thus every noncontextual candidate differs from the target by at least $\nu_d\gamma$ in some retained probability.

Choose the number of repetitions per setting pair as
\begin{align}
K=\left\lceil\frac{8}{\nu_d^2\gamma^2}\log\frac{2d^4}{\delta}\right\rceil.
\label{eq:fixed-identity-allocation-sm}
\end{align}
Substituting into Eq.~\eqref{eq:raw-target-region-sm} gives $r\leq\nu_d\gamma/4$. On the confidence event $p\in\mc R_N$, both $p$ and every candidate $q\in\mc R_N$ lie within $r$ of the empirical frequencies in each retained coordinate. Consequently,
\begin{align}
\max_{j\in\mc J}|q_j-p_j|\leq\max_{j\in\mc J}\bigl(|q_j-f_j|+|f_j-p_j|\bigr)\leq2r\leq\frac{\nu_d\gamma}{2}<\nu_d\gamma.
\end{align}
This contradicts Eq.~\eqref{eq:fixed-identity-probability-gap-sm} for every noncontextual $q$. Hence $\mc R_N$ contains no noncontextual candidate, and the protocol certifies with probability at least $1-\delta$. Since the $d^4$ retained events involve at most $d^4$ distinct setting pairs, the total target-copy count satisfies
\begin{align}
N=|\mc S|K\leq d^4K=O_d\!\left(\gamma^{-2}\log(1/\delta)\right),
\end{align}
proving completeness.
\subsection{Lower bound on the target-copy count}
\label{sec:fixed-lower-sm}

Consider the same devices and supplied identities with the Werner sources $\rho_\gamma$ and $\rho_0$. Let $p_\gamma$ and $p_0$ denote the probabilities that the protocol returns $\mathsf{certify}$ for these two sources. Since $E_{\rm tr}(\rho_\gamma)=\gamma$ and $\rho_0$ is separable, completeness and soundness require $p_\gamma\geq1-\delta$ and $p_0\leq\delta$.

For Bernoulli distributions, write $D_{\mathrm{KL}}(a\Vert b):=a\log\frac{a}{b}+(1-a)\log\frac{1-a}{1-b}$. For a protocol using $N$ copies and local product measurements, relative-entropy data processing inequality~\cite{Lindblad1975} and additivity give
\begin{align}
D_{\mathrm{KL}}(1-\delta\Vert\delta)\leq D_{\mathrm{KL}}(p_\gamma\Vert p_0)\leq N D(\rho_\gamma\Vert\rho_0),
\label{eq:binary-testing-sm}
\end{align}
where $D$ denotes quantum relative entropy. The first inequality follows from $p_\gamma\geq1-\delta>\delta\geq p_0$. (The same information bound also holds for arbitrary joint measurements on the $N$ copies.)

The Werner states share the symmetric and antisymmetric eigenspaces, so
\begin{align}
D(\rho_\gamma\Vert\rho_0)=\frac{1-2\gamma}{2}\log(1-2\gamma)+\frac{1+2\gamma}{2}\log(1+2\gamma)\leq4\gamma^2,
\label{eq:fixed-identity-relative-entropy-sm}
\end{align}
using $\log(1+x)\leq x$. Consequently,
\begin{align}
N\geq\frac{D_{\mathrm{KL}}(1-\delta\Vert\delta)}{4\gamma^2}=\frac{(1-2\delta)\log[(1-\delta)/\delta]}{4\gamma^2}=\Omega \left(\gamma^{-2}\log(1/\delta)\right)
\end{align}
for $0<\delta\leq1/4$. Together with the upper bound, this proves Theorem~\ref{thm:known-identities-sampling}.

\end{document}